\documentclass[EPJ]{webofc}
\usepackage[varg]{txfonts}   
\usepackage{bbm}
\usepackage[colorlinks,citecolor=blue]{hyperref}

\woctitle{The $13^{\rm th}$ International Symposium on Origin of Matter and Evolution of Galaxies}
\begin{document}
\title{Linking neutrino oscillations to the nucleosynthesis of elements}
%
%

\author{Meng-Ru Wu\inst{1} \and
        Gabriel Mart\'inez-Pinedo\inst{1,2} \and
        Yong-Zhong Qian\inst{3}
}


\institute{Institut f{\"u}r Kernphysik (Theoriezentrum), Technische
  Universit{\"a}t Darmstadt, 
  64289 Darmstadt, Germany 
\and
           GSI Helmholtzzentrum f\"ur Schwerioneneforschung,
 64291 Darmstadt, Germany
\and
           School of Physics and Astronomy, University of Minnesota,
  Minneapolis, MN 55455, USA
          }

\abstract{%
  Neutrino interactions with matter play an important role in determining 
  the nucleosynthesis outcome in explosive astrophysical environments such 
  as core-collapse supernovae or mergers of compact objects. 
  In this article, we first discuss our recent work on the importance
  of studying the time evolution of collective neutrino oscillations among
  active flavors in determining their effects on
  nucleosynthesis. We then consider the possible active-sterile neutrino mixing and
  demonstrate the need of a consistent approach
  to evolve neutrino flavor oscillations, matter composition, and the hydrodynamics
  when flavor oscillations can happen very deep inside the supernovae.
}
\maketitle
\section{Introduction}\label{intro}
The formation of the elements in the Universe is
closely related to the weak interactions between neutrinos and matter. 
From the light elements made in the Big-Bang nucleosynthesis to the heavy elements that
are made by the neutron-capture processes in explosive 
astrophysical environments, neutrino interactions 
can interchange protons and neutrons (free or inside the nucleus) 
and play an important role in shaping the neutron-richness 
of the baryonic matter. In the explosive astrophysical
events such as core-collapse supernovae and 
mergers of binary neutron stars or a neutron star with
a black hole, the role of neutrinos in determining 
the property of the ejected matter and the outcome of the
associated nucleosynthesis has been extensively studied,
mostly using hydrodynamical simulations with detailed 
neutrino transport equations 
and post-processing the nucleosynthesis
by an extended nuclear reaction network 
(e.g.,~\citep{Martinez-Pinedo:2013jna,Just:2014fka}).
 
However, an important aspect that currently cannot
be modelled in the hydrodynamical simulations is the quantum
phenomenon of neutrino flavor oscillations.
As neutrinos with different flavors interact differently with 
matter, any mechanism that alters the flavor
of the neutrinos after their production can potentially
affect the prediction of the matter property and the
outcome of the nucleosynthesis (e.g.,~\citep{Duan:2010af,MartinezPinedo:2011br}).

Neutrino flavor oscillations which arise from the mixing
between their flavor eigenstates and the mass eigenstates have
successfully accounted for the results of terrestrial and Solar neutrino experiments.
In fact, nearly all the mixing parameters are precisely measured, except for the 
sign of the atmospheric mass-squared difference and the CP violating
phase(s) \citep{Agashe:2014kda}. 
However, in the extreme astrophysical environments, 
due to the high baryonic density which can be
higher than the nuclear saturation density and the high temperature,
large neutrino number density which is comparable to the densities of baryons and
electrons is typically present. 
As a result, neutrino self-interactions that must be considered in 
modelling their flavor oscillations lead to a non-linear coupling between
the different neutrino quantum states. 
Despite numerous works studying neutrino flavor oscillations
in those environments during the past decade 
(see e.g.,~\citep{Duan:2010bg,Duan:2015cqa} for reviews and the references therein),
it remains a challenging and exciting problem to be solved
in order to fully appreciate the role of neutrinos in supernova explosions
and in the nucleosynthesis of elements. In this article, we discuss some of 
our recent works along this direction in improving
the link between neutrino flavor oscillations and
the nucleosynthesis of elements, particularly in core-collapse supernovae.

\section{Neutrino flavor oscillations in medium}\label{sec-form}
For neutrino flavor oscillations in the dilute gas limit such that
neutrinos kinematically decouple from matter,
the equation of motion for the neutrino density matrix 
$\varrho(t,\mathbf{x},\mathbf{p})$ is given by \citep{Vlasenko:2013fja}
\begin{equation}\label{eq-eomfull}
\frac{\partial\varrho(t,\mathbf{x},\mathbf{p})}{\partial t}+
\mathbf{\hat v}\cdot\nabla\varrho(t,\mathbf{x},\mathbf{p})
=-i[H(t,\mathbf{x},\mathbf{p}),\varrho(t,\mathbf{x},\mathbf{p})].
\end{equation}
The Wigner-transformed density matrix $\varrho(t,\mathbf{x},\mathbf{p})$ can be explicitly written
in the flavor basis for the active neutrinos:
\begin{equation}\label{eq-dmtr}
\varrho(t,\mathbf{x},\mathbf{p})=
\left [
\begin{array}{ccc}
\varrho_{ee} & \varrho_{e\mu} & \varrho_{e\tau} \\
\varrho_{e\mu}^* & \varrho_{\mu\mu} & \varrho_{\mu\tau} \\
\varrho_{e\tau}^* & \varrho_{\mu\tau}^* & \varrho_{\tau\tau} \\
\end{array}
\right ],
\end{equation}
where the diagonal terms $\varrho_{\alpha\alpha}(t,\mathbf{x},\mathbf{p})
=f_{\nu_\alpha}(t,\mathbf{x},\mathbf{p})$ are the statistical phase-space
distribution functions of neutrinos with flavor $\alpha$.
The off-diagonal (correlation) terms encode the information of neutrino flavor mixing.
The Hamiltonian $H(t,\mathbf{x},\mathbf{p})=H_{\rm vac}(p)+H_m(t,\mathbf{x})+H_{\nu\nu}(t,\mathbf{x},\mathbf{p})$
contains the contribution from the vacuum neutrino mixing, neutrino forward-scattering
with matter \citep{Mikheev:1986gs,Wolfenstein}, and neutrino forward-scattering among themselves
\citep{Fuller:1987,Pantaleone:1992eq,Sigl:1992fn}.
$H_{\rm vac}(p)=UM^2U^\dagger/2p$ where $U$ is the unitary mixing matrix, 
$M={\rm diag}(m_1,m_2,m_3)$ with $m_i$ being the mass of the $i$th neutrino
mass eigenstate. $H_m(t,\mathbf{x})=\sqrt{2}G_F
[n_e(t,\mathbf{x}){\rm diag}(1,0,0)-n_n(t,\mathbf{x})\mathbbm{I}_{3\times 3}/2]$,
$n_e(t,\mathbf{x})$ is the net electron number density
and $n_n(t,\mathbf{x})$ is the neutron number density. 
The $\nu$-$\nu$ Hamiltonian
\begin{equation}
H_{\nu\nu}(t,\mathbf{x},\mathbf{p})=
\frac{\sqrt{2}G_F}{(2\pi)^3}\int d^3q (1-\mathbf{\hat p}\cdot\mathbf{\hat q})
\lbrace\varrho(t,\mathbf{x},\mathbf{q})-\bar\varrho(t,\mathbf{x},\mathbf{q})+
{\rm Tr}[\varrho(t,\mathbf{x},\mathbf{q})-\bar\varrho(t,\mathbf{x},\mathbf{q})]\mathbbm{I}_{3\times 3}\rbrace,
\end{equation}
where $\bar\varrho$ is the density matrix for antineutrinos defined
in the same way as in Eq.~(\ref{eq-dmtr}).
In the above equations, we have neglected the sub-leading terms 
in the Hamiltonian which 
can cause the helicity coherence \citep{Vlasenko:2013fja} and 
the beyond-mean-field correlations \citep{Volpe:2013uxl}.

The above formulation can be easily generalized to describe the
flavor mixing between active neutrinos and the sterile neutrinos ($\nu_s$)
by enlarging $\varrho$ to include the components $\varrho_{\alpha s}$
that however do not contribute to the Hamiltonian $H_{\nu\nu}$
in the leading order. Therefore, the only change to the Hamiltonian
is the addition of vacuum mixing entries in $H_{\rm vac}$.

\section{Collective neutrino oscillations and supernova nucleosynthesis}\label{sec-cno}
To apply the above formalism in the astrophysical environments
such as supernovae, we first assume that all neutrinos decouple
from matter kinematically at a sharp neutrinosphere $r=R$.
We further make the assumptions that the supernova environment is 
spatially spherically-symmetric, temporally stationary during the
time of neutrino propagation, and the flavor evolution of neutrinos 
preserves these symmetries. In this case, Eq.~(\ref{eq-eomfull}) can
be reduced to :
\begin{equation}\label{eq-eomred}
\frac{d\varrho(t,r,u,E)}{dr}=-i[H(t,r,u,E),\varrho(t,r,u,E)],
\end{equation}
where $E$ is the neutrino energy, $u=\cos\theta_{\rm em}$
with $\theta_{\rm em}$ being the emission angle of the neutrinos
from the neutrinosphere w.r.t to the radial direction.
The initial conditions are given by setting the non-zero diagonal elements
of $\varrho(t,r,u,E)$ equal to the neutrino distribution 
function $f_{\nu_\alpha}(t,r,u,E)$ which can be parametrized
or given by supernova simulations with detailed neutrino transport.
Supplied with the density profiles $n_e(t,r)$ and $n_n(t,r)$, 
Eq.~(\ref{eq-eomred}) can then be solved for each given $t$.

\begin{figure}[t]
\centering
\includegraphics[width=0.28\columnwidth, angle=-90]{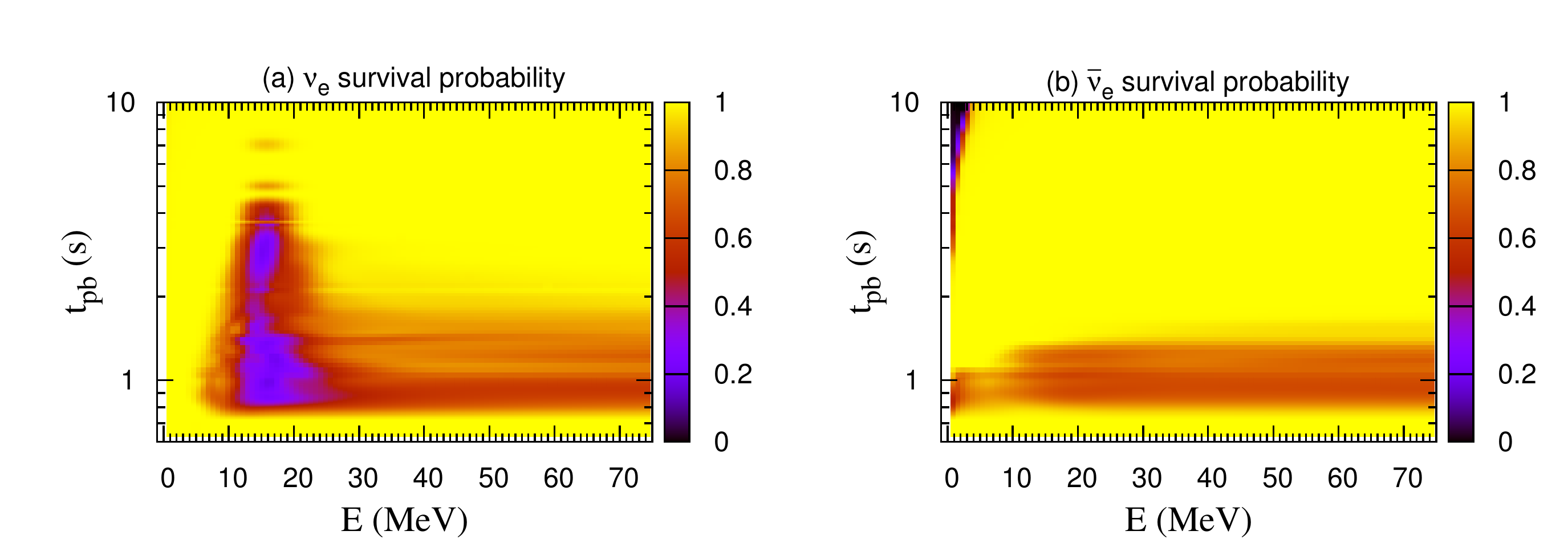}
\caption{The angle-averaged survival probabilities of
(a) $\nu_e$ and (b) $\bar\nu_e$ as functions of 
the neutrino energy $E$ and the post-bounce time $t_{\rm pb}$.
(Reprinted figure from \citep{Wu:2014kaa}; 
\href{http://dx.doi.org/10.1103/PhysRevD.91.065016}{copyright (2015) by the American
Physical Society.})}
\label{fig-1}
\end{figure}

We performed a set of comprehensive numerical calculations 
to map out the flavor conversion probabilities 
$P_{\nu_\alpha\rightarrow\nu_\beta}(t,r,u,E)$ between active flavors \citep{Wu:2014kaa}
for the supernova models of \citep{Fischer:2009af}. We found that within the first $\sim 500$~km,
the neutrino flavor conversion is dominated by the $\nu$-$\nu$
contribution. This leads to the so-called 
``collective neutrino oscillations'' \citep{Duan:2010bg,Duan:2015cqa},
which may give rise to a sharp transition of $P_{\nu_\alpha\rightarrow\nu_\beta}(t,r,u,E)$
at some specific $E$ (the ``spectral splits/swaps'').
Fig.~\ref{fig-1} shows the angle-averaged survival probabilities for the
initial $\nu_e$ and $\bar\nu_e$ as functions of $E$ and 
the time post supernova core-bounce, $t_{\rm pb}$, 
at $r=500$~km where the collective neutrino 
oscillations have ceased for the $18$~$M_\odot$ supernova. 
We see that the feature of the spectral splits
is indeed present in our calculations. More importantly, the
survival probabilities change substantially as the supernova evolves
with time. At the later stage, the flavor conversions 
between $\nu_e\leftrightarrow\nu_{\mu,\tau}$ and
$\bar\nu_e\leftrightarrow\bar\nu_{\mu,\tau}$
are more suppressed. 

We stress here that this time evolution has important consequences for 
supernova nucleosynthesis in both the $\nu$-driven wind and 
the $\nu$(-induced) nucleosynthesis in the supernova envelopes. 
For the $\nu$-driven wind which may be a site
of heavy element formation, the time scale relevant for the 
neutrino-matter interactions to change the ejecta composition can
last a few seconds if it is initially proton-rich such that
the $\nu$p process can occur \citep{Frohlich:2005ys,Pruet:2005qd}. 
Similarly, for the $\nu$(-induced)
nucleosynthesis (e.g.~\citep{Heger:2003mm,Sieverding:2015oya}), 
it is the total exposure of nuclei to
the neutrino fluence that determines the final yields. In this study, we found 
that due to the suppression of flavor conversion in
the $\bar\nu$ sector, collective neutrino flavor oscillations
have little impact on the $\nu$p process.
As substantial flavor conversion occurs in the 
$\nu$ sector, the production of rare nuclei such as 
$^{138}$La and $^{180}$Ta may be enhanced by
the flavor conversion of $\nu_e\leftrightarrow\nu_{\mu,\tau}$. 
Nevertheless, we note that these results
are subject to change once the effect of symmetry breaking
of neutrino flavor oscillations 
(see e.g.,~\cite{Raffelt:2013rqa,Chakraborty:2015tfa,Abbar:2015mca,Abbar:2015fwa,Dasgupta:2015iia})
can be self-consistently taken into account in the future.

\section{The interplay between flavor oscillations and hydrodynamics}\label{sec-oschy}
The discussion in the last section is based on the assumption that 
neutrino flavor oscillations do not change the hydrodynamic
variables such as the baryonic density $\rho$, the temperature
$T$, the fluid velocity $v$, and the matter composition. 
However, if neutrino flavor oscillations
happen deep enough inside supernovae where the neutrino-matter
interactions are still important in setting up the hydrodynamic
properties of the environment, one has to evolve the flavor
equations along with the hydrodynamic equations. In principle,
for a fully consistent derivation, one needs to extend Eq.~(\ref{eq-eomfull})
to full quantum kinetic equations \citep{Vlasenko:2013fja} by
including the collision terms
and couple them with the hydrodynamic equations. However, as a 
first step, we make the assumption that the neutrino interactions
with matter are weak enough such that the collision terms can
still be neglected for flavor evolution while those interactions may be strong enough
to change the hydrodynamic properties.

Based on the above, we have coupled the further reduced flavor 
evolution equation using the so-called ``single-angle approximation'' \citep{Duan:2010bg}
which assumes that the flavor evolution history is independent
of the neutrino emission angles:
\begin{equation}\label{eq-eomsa}
\frac{d\varrho(r,E)}{dr}=-i[H(r,E),\varrho(r,E)],
\end{equation}
with the steady-state hydrodynamic equations that may adequately 
describe the physical conditions in the $\nu$-driven wind 
from the proto-neutron star (PNS) \citep{Qian:1996xt,Thompson:2001ys}:
\begin{subequations}\label{eq-hydro}
\begin{align}
& \dot{M}=4\pi r^2\rho vy, \\
& \frac{1}{y}\frac{dy}{dr}+\frac{1}{\varepsilon +P}\frac{dP}{dr}=0, \\
& \frac{d\varepsilon}{dr}-\frac{\varepsilon +P}{\rho}\frac{d\rho}{dr}
-\rho\frac{{\dot q}_\nu}{vy}=0,
\end{align}
\end{subequations}
where $\dot M$ is the constant mass outflow rate of the ejecta, 
$v$ is its radial velocity, $y^2=(1-2GM/r)/(1-v^2)$,
$G$ is the Newtonian gravitational constant,
$M$ is the mass of the PNS,
$\varepsilon$ is the total energy density, $P$ is the pressure,
and ${\dot q}_\nu$ is the net energy gain/loss rate per unit 
mass by $\nu$ heating and cooling. For ${\dot q}_\nu$,
we have included the charged-current $\nu$ absorption,
$\nu\bar\nu$ annihilation and their reverse reactions,
and $\nu$ scattering with $e^\pm$ and nucleons.
Detail expressions will be reported in a forthcoming
publication \citep{Wu:2015}.

For the matter composition of the wind, we consider the phase when 
the temperature is still high enough so that matter consists of free protons, neutrons,
and $e^{\pm}$.
In this case, the matter composition is determined by 
the electron number fraction $Y_e=n_e/(\rho/m_u)$ where
$m_u$ is the atomic mass unit
and the evolution of $Y_e$ is governed by
\begin{equation}\label{eq-Ye}
(vy)\frac{dY_e}{dr}=(\lambda_{\nu_en}+\lambda_{e^+n})(1-Y_e)+(\lambda_{\bar\nu_ep}+\lambda_{e^-p})Y_e,
\end{equation}
where $\lambda_{\nu_en}$, $\lambda_{e^+n}$, $\lambda_{\bar\nu_ep}$ and $\lambda_{e^-p}$ 
are the corresponding charged-current reaction rates.

We consider here in particular the reduced flavor subspace of $\nu_e$ ($\bar\nu_e$) and 
$\nu_s$ ($\bar\nu_s$), with a mass-squared difference
$\delta m^2\approx 1.75$~eV$^2$ and the vacuum mixing
angle corresponding to $\sin^22\theta\approx 0.1$, as indicated by the 
neutrino anomalies \citep{Acero:2007su,Mention:2011rk}.
Such an active-sterile flavor conversion
for the initial $\nu_e$ and $\bar\nu_e$ can happen at $Y_e\approx 1/3$ such 
that both ${\dot q}_\nu$ and $Y_e$ are 
greatly influenced as suggested by previous studies without considering the feedback
of flavor oscillations on hydrodynamics \citep{McLaughlin:1999pd,Tamborra:2011is,Wu:2013gxa}. 

Eqs.~(\ref{eq-eomsa})--(\ref{eq-Ye}) can be solved given the boundary conditions below:
\begin{itemize}
\item The $\nu$ luminosity and the temperature, 
$L_{\nu_\alpha}$ and $T_{\nu_\alpha}$, at the PNS surface $r=R$, 
assumed to be given by a Fermi-Dirac distribution with zero chemical potential.
This specifies the neutrino density matrix element
$\varrho_{\alpha\alpha}(R,E)=f_{\nu_\alpha}(E)=1/(1+e^{E/T_{\nu_\alpha}})$.
\item The hydrodynamic conditions at the PNS surface, $T(R)=T_{\nu_e}$, $\dot{q}(R)=0$ and ${\dot Y}_e(R)=0$.
\item An outer boundary temperature $T_b$ at some large radius $r\gg R$.
\end{itemize}
We have performed such a calculation with the parameters
$(L_{\nu_e},L_{\nu_e},L_{\nu_{\mu,\tau}})=(1.67,2.01,2.58)\times 10^{51}$~erg/s,
$(T_{\nu_e},T_{\nu_e},T_{\nu_{\mu,\tau}})=(2.68,3.78,3.71)$~MeV,
$M=1.282$~$M_\odot$, $R=18.07$~km, and $T_b=0.12$~MeV at $r=10^3$~km.
Those values are taken with the guide of a supernova simulation \citep{Martinez-Pinedo:2013jna}.
To understand the role of the convoluted feedback of the composition and hydrodynamic changes
on the flavor oscillations, we have performed additional calculations as follows:
\begin{enumerate}
\item No feedback: Eq.~(\ref{eq-eomsa}) decoupled from Eq.~(\ref{eq-hydro}) and Eq.~(\ref{eq-Ye}), 
i.e., we first derive $Y_e(r)$ and $\rho(r)$ from Eq.~(\ref{eq-hydro}) and Eq.~(\ref{eq-Ye}) 
without considering flavor oscillations. 
We then evolve Eq.~(\ref{eq-eomsa}) with the derived $Y_e(r)$ and $\rho(r)$.
\item Feedback on $Y_e$ only: Eq.~(\ref{eq-eomsa}) coupled with Eq.~(\ref{eq-Ye}) but decoupled from Eq.~(\ref{eq-hydro}),
i.e., we first derive $\rho(r)$ from Eq.~(\ref{eq-hydro}) and Eq.~(\ref{eq-Ye}) 
without considering flavor oscillations.
We then evolve Eq.~(\ref{eq-eomsa}) and Eq.~(\ref{eq-Ye}) with the derived $\rho(r)$.
\end{enumerate}

\begin{figure}[t]
\centering
\includegraphics[width=0.76\columnwidth]{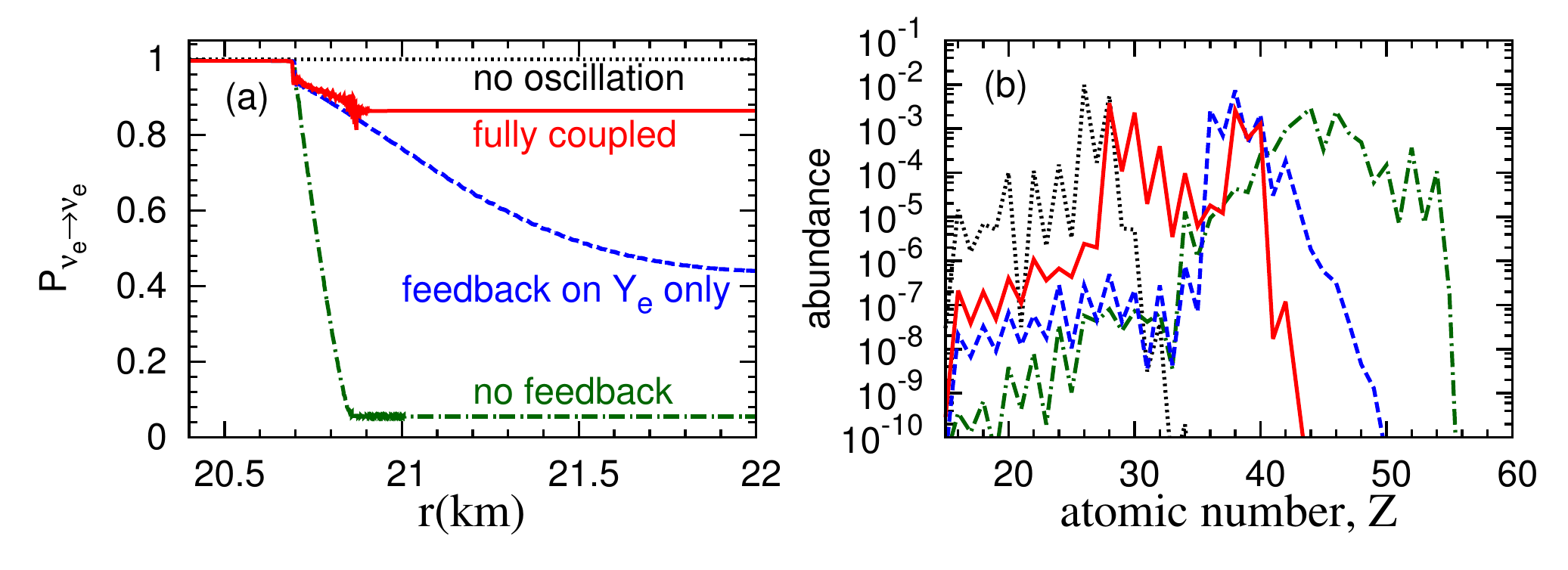}
\caption{(a) The average $\nu_e$ survival probability
as a function of radius for different cases discussed in the text, and
(b) the corresponding elemental abundances of nucleosynthesis as a function of atomic number.}
\label{fig-2}
\end{figure}

In Fig.~\ref{fig-2}(a), we show the 
energy-averaged $\nu_e$ survival probability $P_{\nu_e\rightarrow\nu_e}$
as a function of radius for all three cases described above.
It clearly shows that both $Y_e$ and hydrodynamic evolution
have strong impact on the neutrino flavor conversion.
In the case of ``no feedback'', nearly all the initial $\nu_e$ are 
converted to $\nu_s$, due to the so-called ``matter-neutrino resonance'' (MNR)
mechanism \citep{Malkus:2014iqa,Wu:2015fga}. For the ``feedback on $Y_e$ only'' case, 
$\nu_e$ are less converted to $\nu_s$ but the flavor conversion
process takes place up to a much larger radius.
As for the fully coupled case, only $\sim 15\%$ of $\nu_e$
are converted. The main differences between those results
arise mainly because MNR can be extended to a much larger radius
once $Y_e$ is affected by $\nu$ interactions.
However, this extended resonance may not remain stable,
depending on the velocity of the ejecta, which is affected by
the change of ${\dot q}_\nu$. The details will be reported in
a forthcoming publication \citep{Wu:2015}.

In Fig.~\ref{fig-2}(b), we further show the resulting elemental abundances
as a function of the atomic number $Z$. Because $P_{\nu_e\rightarrow\nu_e}$ differs 
when a different level of coupling 
among Eqs.~(\ref{eq-eomsa})--(\ref{eq-Ye}) is employed,
the nucleosynthesis outcome can be dramatically different.
As expected, when $\nu_e$ are converted more to $\nu_s$,
the matter composition becomes more neutron-rich,
which results in the production of heavier elements.

\section{Summary}\label{sec-summary}
In this article, we have discussed two aspects in
connecting neutrino flavor oscillations to the nucleosynthesis
of elements. We have shown in Sec.~\ref{sec-cno} that due to the time evolution
of the neutrino characteristics and the density structure
above the PNS in supernovae, it is important to include
this time-dependence when studying the collective neutrino 
flavor oscillations and their impact on the nucleosynthesis
in the $\nu$-driven wind and in the supernova envelopes.
In Sec.~\ref{sec-oschy}, we have shown that when
neutrino flavor oscillations take place very close to 
the PNS, which may happen when considering the possible
active-sterile neutrino flavor transformation, 
the change of the matter composition and hydrodynamic
quantities may have a large impact on the $\nu$ 
flavor evolution histories, thereby complicating the determination of 
the nucleosynthesis outcome in the
$\nu$-driven wind.\\

\textbf{Acknowledgements:} This work was partly supported by the Helmholtz Association (HGF) 
through the Nuclear Astrophysics Virtual Institute (VH-VI-417),
the Helmholtz International Center for FAIR
within the framework of the LOEWE program launched
by the state of Hesse
and by the US DOE grant DE-FG02-87ER40328 at UMN.

\bibliography{omegproc_arXiv}

\end{document}